\documentclass[aps,pra,superscriptaddress,showpacs]{revtex4} 

\usepackage{dcolumn}  
\usepackage{graphicx}  
\usepackage{epsf}  
\usepackage{epsfig}  
\usepackage{amsmath}  
\usepackage{bm}  
\usepackage{times}  
\usepackage{nicefrac}  
\usepackage{color}  

\newcolumntype{.}{D{x}{}{-1}}  
  
\def\corresponds{{\lower.2ex\hbox{=}}{\rm\kern-.75em^\triangle}}  
\def\succsim{\succ\kern-.9em_\sim\kern.3em}  
\def\precsim{\prec\kern-1em_\sim\kern.3em}  
\def\slantfrac#1#2{\kern1em^{#1}\kern-.3em/\kern-.1em_{#2}}  
\def\lfrac#1#2{{}^{#1\!}\kern-.0em/_{#2}}  
  
\def\buildrel#1\under#2{\mathrel{\mathop{\kern0pt #2}\limits_{#1}}}

\def\mr#1{{\mathrm{#1}}}

\begin{document}  
\preprint{Version 1.0}  
 
\title{Magnetic particle hyperthermia: \\ 
N\'eel relaxation in magnetic nanoparticles under circularly 
polarized field} 

\author{P. F. de Ch\^atel}  
\affiliation{Institute of Nuclear Research, P.O.Box 51,  
H-4001 Debrecen, Hungary}
\affiliation{Institute of Metal Research, 
Chinese Academy of Sciences, Shenyang 110016, P.R. China}  
 
\author{I. N\'andori}  
\affiliation{Institute of Nuclear Research, P.O.Box 51,  
H-4001 Debrecen, Hungary}  

\author{J. Hakl}  
\affiliation{Institute of Nuclear Research, P.O.Box 51,  
H-4001 Debrecen, Hungary}  
  
\author{S. M\'esz\'aros}  
\affiliation{Institute of Nuclear Research, P.O.Box 51,  
H-4001 Debrecen, Hungary}  
 
\author{K. Vad}  
\affiliation{Institute of Nuclear Research, P.O.Box 51,  
H-4001 Debrecen, Hungary}  
 
\begin{abstract}  
The mechanism of magnetization reversal in single-domain ferromagnetic 
particles is of interest in many applications, in most of which losses 
must be minimized. In cancer therapy by hyperthermia the opposite 
requirement prevails: the specific loss power should be maximized. Of 
the mechanisms of dissipation, here we study the effect of N\'eel 
relaxation on magnetic nanoparticles unable to move or rotate and compare 
the losses in linearly and  circularly polarized field. We present exact 
analytical solutions of the Landau--Lifshitz equation as derived from the 
Gilbert equation and use the calculated time-dependent magnetizations to 
find the energy loss per cycle. In frequencies lower than the Larmor 
frequency linear polarization is found to be the better source of heat 
power, at high frequencies (beyond the Larmor frequency) circular 
polarization is preferable.
\end{abstract}  
  
\pacs{05.40.Jc, 05.10.Gg, 83.10.Mj}

\maketitle  
 
%---------------------------------------------------------------
% INTRODUCTION 
%---------------------------------------------------------------
\section{Introduction} 
\label{intro}

The dynamics of the magnetization of single-domain ferromagnetic bodies 
under various conditions has been studied with widely different motivations. 
Ferromagnetic resonance has provided perhaps the best known example. In 
this case a strong static magnetic field {\bf B} generates and stabilizes the 
single-domain state. The torque ${\bf  G} = {\bf\mu \times B}$ exercised 
by this field, being perpendicular to ${\bf B}$, causes a precession of 
the magnetic moment ${\bf \mu}$ around the field, leaving the angle between 
${\bf\mu}$ and ${\bf B}$ intact. The resonance is detected by means of an 
alternating field perpendicular to ${\bf B}$, whose energy is absorbed at a 
maximum rate at the Larmor frequency, $\omega_L = \gamma \vert{\bf B}\vert$. 

Without a strong static field, in a sizable sample the single-domain 
state can be only achieved in a material of extremely large 
permeability. This requirement dictated the choice of ``soft iron'' 
in the famous Einstein - de Haas experiment \cite{EiHa1915}, which 
provided the first experimental determination of the gyromagnetic ratio 
$\gamma$. The limitation to soft magnetic materials can be circumvented 
though with very small samples. Clearly, samples smaller than the 
domain-wall thickness cannot contain more than one domain. 

The dynamics of the magnetic moment of nanometre-sized single-domain 
particles is of interest in connection with a number of applications. 
In magnetic recording, the processes taking place in magnetized particles 
under minor perturbations (reading) and during magnetization reversal 
(switching) are of paramount interest. In this case, the nanoparticles 
are immobile, their place and orientation are fixed. A fast process of 
magnetization reversal with minimal losses is the most important requirement 
for applicability in magnetic recording. 

Ferrofluids contain mobile 
magnetic nanoparticles, whose position and orientation can be controlled 
by a magnetic field. In their application in hyperthermia, the most 
important requirement is the absorption of large losses under repeated 
magnetization reversal. More precisely, the figure of merit to be maximized 
is the specific absorption rate (also called the specific loss rate, or 
heating rate), i.e., the amount of energy absorbed per second by a unit 
mass of magnetic nanoparticles. The conditions under which maximization 
has to be realized are quite demanding, as they are dictated by biophysical 
factors \cite{Hergt}.

The mobility of magnetic particles in ferrofluids is an important feature 
in the context of losses under repeated magnetization reversal. As the 
particles are free to rotate, the potential energy $E=-{\bf \mu\cdot B}$ 
can be minimized in two ways: either the magnetic moment rotates within the 
particle until it is aligned with the field, or the particle rotates as 
a whole. Two different processes are involved, N\'eel and Brown relaxation, 
respectively. In the latter the energy is transferred directly to the fluid, 
due to the viscous flow arising at the surface of the particle. In N\'eel 
relaxation the energy is transferred to the lattice, i.e., it is absorbed 
by phonons and magnons, before ordinary heat transport conveys it to the 
fluid. Both processes are characterized by appropriate relaxation times, 
$\tau_N$ and $\tau_B$, a parameterization that conceals the difference 
between the two mechanisms. The apparent similarity is further emphasized 
by the term 'magnetic viscosity', a misleading synonym for N\'eel relaxation.

The common practice in hyperthermia is to expose magnetic nanoparticles 
present in a ferrofluid, preferably inside the malignant tumours, to a 
magnetic field alternating at a frequency of the order of $10^5$ Hz. The 
optimization of loss energy with respect to the amplitude and frequency 
of this linearly polarized field has been studied in detail \cite{CoFa2002}. 
However, the effect of circularly polarized field received relatively little
attention. This neglect is not justified though, as there is no obvious 
reason why the specific absorption rate should be indifferent to the 
nature of polarization.

In the present paper we report on the first stage of a systematic 
theoretical study of the dynamic response of a system of magnetic 
nanoparticles to  linearly and circularly polarized magnetic field. 
The ultimate goal of this study is an analysis of the criteria for a 
maximum output of heating power applicable in hyperthermia. The relative 
importance of N\'eel and Brown relaxations is not clear. Hergst et al. 
\cite{Hergt} have made susceptibility measurements on a colloidal 
suspension of maghemite ($\gamma$Fe$_2$O$_3$) particles at frequencies 
ranging from 10 Hz and 1 MHz. The very broad peak in the imaginary part 
of the susceptibility found at 1 kHz disappeared when the particles were 
made immobile by letting the solvent freeze. In this case, viscous 
losses (Brownian relaxation) is clearly dominant. On the other hand, 
these authors also pointed out that the mobility of particles trapped 
in tumour tissue is not known. If they are clogged to the extent of 
not being able to rotate, the N\'eel mechanism is the only relaxation 
process available. 
Here we present analytic results for an immobilized single-domain particle. 
In Section 2 we briefly summarize the theoretical tools available for 
the study of the N\'eel mechanism and recapitulate the relation between 
the Landau--Lifshitz and Gilbert equations. Section 3 treats the 
case of linearly polarized magnetic fields, for which we present exact 
analytical solutions of the Landau--Lifshitz equation, which are 
used to determine the heating power as a function of frequency and 
amplitude of the alternating field. In Section 4 approximate analytic 
solutions are presented for circular polarization, which are exact under 
specific initial conditions. For arbitrary initial conditions, those 
solutions give an exact description of the dynamics in the steady state, 
which is stabilized after a short initial transient time-interval. The 
heating power is calculated for this steady state.

%-------------------------------------------------------------- 
% Neel relaxation 
%--------------------------------------------------------------
\section{N\'eel relaxation} 
\label{neel} 

We shall describe the magnetization processes in an isotropic 
single-domain particle by means of the Landau--Lifshitz equation 
\cite{LaLi1935}, which is generally used to consider the N\'eel 
relaxation mechanism \cite{Neel} of magnetic nanoparticle systems. It 
is customary to write the Landau--Lifshitz equation in the form
\begin{equation}
\label{LL}
\frac{\mr{d}}{\mr{d}t} {\bf m} = 
\mu_0 \gamma \left([{\bf m\times H}] 
+ \frac{\alpha}{m_S} [[{\bf m\times H]\times m}]\right),
\end{equation}
where ${\bf m} = {\bf\mu}/V$, V is the volume of the particle,  
$\mu_0= 4\pi \times 10^{-7}$ N/A$^2$ is the permeability of free space,  
$\gamma$ is an effective gyromagnetic ratio, $\alpha$ is a dimensionless 
damping constant, $m_S$ is the saturation magnetization and 
${\bf H} = \mu_0^{-1} {\bf B}$ is the applied magnetic field. Formally, 
the Gilbert equation \cite{Gilbert},
\begin{equation}
\label{G}
\frac{\mr{d}}{\mr{d}t} {\bf m} = \mu_0 \gamma_0 
\left([{\bf m \times H}] 
- \eta [{\bf m} \times \frac{\mr{d}}{\mr{d}t} {\bf m}]\right)
\end{equation}
is equivalent to the Landau-Lifshitz equation, as it can be verified 
\cite{Gilbert} with the identifications $\alpha=\mu_0 \gamma_0 \eta m_S$ 
and $ \gamma = \gamma_0/(1+\alpha^2)$ of the parameters of the latter 
in terms of those of the former. Here $\gamma_0=8.82 \cdot 10^{10}$ 
Am$^2$/Js is the gyromagnetic ratio of the electron and $\eta$ is the 
damping constant. As it was pointed out by Kikuchi \cite{Ki1956}, the 
factor $1/(1 + \alpha^2)$ appearing on the right-hand side of the 
Landau--Lifshitz equation as derived from the Gilbert equation 
removes a ``physically implausible situation'' \cite{Ma1987} in the 
predicted behaviour at large damping.

The most conspicuous feature of the Landau--Lifshitz equation is that 
the vector standing on the right-hand side is perpendicular to the 
magnetization vector. It follows then that 
$dm^2/dt = {\bf m} \cdot d{\bf m}/dt = 0$, that is, the magnetization 
vector's magnitude does not change under the influence of the external 
field. Therefore, it is convenient to rewrite equation (\ref{LL}) in 
terms of the unit vector ${\bf M} = {\bf m}/m_S$:
\begin{equation}
\label{LLG}
\frac{\mr{d}}{\mr{d}t} {\bf M} = \gamma' [{\bf M \times H}] 
- \alpha' [[{\bf M\times H]\times M}].
\end{equation}
The new coefficients are given in terms of the ones in equations (\ref{LL}) 
and (\ref{G}) as $\gamma' = \mu_0 \gamma = \mu_0 \gamma_0/(1+\alpha^2)$
and $\alpha' = \mu_0 \gamma \alpha = \mu_0^2 \gamma_0^2 \eta m_S/(1+\alpha^2)$. 
With these definitions of the coefficients, equation (\ref{LLG}) is the 
Landau--Lifshitz equation as derived from the Gilbert equation \cite{Br1979}, 
which will be referred to as the Landau--Lifshitz--Gilbert (LLG) equation in 
what follows.

%-------------------------------------------------------------- 
% Linearly polarized applied field
%--------------------------------------------------------------
\section{Linearly polarized applied field} 
\label{lin} 

In this section, we present an exact analytic solution for the LLG 
equation (\ref{LLG}) for an isotropic magnetic particle subjected to a 
linearly polarized, i.e. alternating applied magnetic field. The results 
enable the numerical evaluation of the heating rate per particle. The 
alternating field is assumed to be applied in the $x$-direction, 
\begin{eqnarray}
\label{osc_field}
{\bf H} &=& \frac{\omega_L}{\gamma'} \, \, (\cos(\omega t),0,0),
%\nonumber \\
%{\bf H} &=& \frac{\alpha_N}{\alpha'} \, \, (\cos(\omega t),0,0),
\end{eqnarray}
where $\omega_L = \gamma' \vert{\bf H}\vert$ is the Larmor frequency,
$\omega$ is angular frequency of the alternating external magnetic field.

%------------------------------------------------------------
\subsection{Free precession}
%------------------------------------------------------------

In the case of vanishing relaxation ($\alpha' = 0$), the LLG equations 
for the Cartesian components of the magnetization vector are 
\begin{eqnarray}
\label{osc_eq_nonrelax}
\frac{\mr{d}}{\mr{d}t} M_x &=&  0,
\nonumber \\
\frac{\mr{d}}{\mr{d}t} M_y &=&  \cos(\omega t) \, \omega_L M_z,
\nonumber \\
\frac{\mr{d}}{\mr{d}t} M_z &=& -\cos(\omega t) \, \omega_L M_y.
\end{eqnarray}
The analytic solution of these coupled equations for all possible initial 
conditions can be written as
\begin{eqnarray}
\label{osc_sol_norelax}
M_{x}(t) &=& M_{x0},
\nonumber \\
M_{y}(t) &=& \sqrt{1-M_{x0}^{2}} \, \,
\sin\left[\frac{\omega_L}{\omega}\sin(\omega t) + \delta_0\right],
\nonumber \\
M_{z}(t) &=& \sqrt{1-M_{x0}^{2}} \, \,
\cos\left[\frac{\omega_L}{\omega}\sin(\omega t) + \delta_0\right],
\end{eqnarray}
where $M_{x0}$ and $\delta_0$ are determined by the initial conditions:  
$M_{x0} = M_x(0)$ and $\delta_0 = \tan^{-1}(M_{y}(0)/M_{z}(0))$. The 
interpretation of equation (\ref{osc_sol_norelax}) is that the unit 
vector representing the magnetization is precessing around the $x$ axis 
with a time-dependent angular velocity. At time $t$, its projection on 
the $yz$ plane is at an angle $\delta(t) = \tan^{-1}(M_{y}(t)/M_{z}(t))= 
[(\omega_L /\omega)\sin(\omega t) + \delta_0]$ from the $z$ axis. The 
angular velocity of the precession is changing periodically. If 
$(\omega_L /\omega) >> 1$, precession in a given sense prevails for 
many full cycles, before it slows down and changes sign. For the 
static limit, i.e., $\omega \to 0$, the solution (\ref{osc_sol_norelax}) 
reduces to 
\begin{eqnarray}
\label{stat_sol_norelax}
M_{x}(t) &=& M_{x0},
\nonumber \\
M_{y}(t) &=& \sqrt{1-M_{x0}^{2}} \, \,
\sin\left[\omega_L t + \delta_0\right],
\nonumber \\
M_{z}(t) &=& \sqrt{1-M_{x0}^{2}} \, \,
\cos\left[\omega_L t + \delta_0\right].
\end{eqnarray}
This is the limit when the period of the alternating field is so long 
that the Larmor precession seems to proceed uninterrupted. In the opposite 
limit, $(\omega_L /\omega) << 1$, there is only an oscillation of small 
amplitude about $\delta_0$ (i.e., $\delta(t)\approx\delta_0$).

%------------------------------------------------------------
\subsection{Damped precession}
%------------------------------------------------------------

In the presence of N\'eel relaxation the LLG equations for the 
Cartesian components of the magnetization for linearly polarized
applied field are
\begin{eqnarray}
\label{osc_eq}
\frac{\mr{d}}{\mr{d}t} M_x &=& \alpha_N \cos(\omega t) \,\, 
(1-M_x^2),
\nonumber \\
\frac{\mr{d}}{\mr{d}t} M_y &=& \cos(\omega t) \,
\left( \omega_L M_z - \alpha_N M_x M_y \right),
\nonumber \\
\frac{\mr{d}}{\mr{d}t} M_z &=& -\cos(\omega t) \,
\left( \omega_L M_y + \alpha_N M_x M_z \right),
\end{eqnarray}
where we have introduced $\alpha_N = \alpha' \vert{\bf H}\vert$  as a 
measure of damping. The analytic solution is
\begin{eqnarray}
\label{osc_sol_relax}
M_{x}(t) &=& \frac{(M_{x0}-1)+(M_{x0}+1)
\exp{\left[\frac{2\alpha_N}{\omega}\sin(\omega t)\right]}} 
{(1-M_{x0})+(M_{x0}+1)
\exp{\left[\frac{2\alpha_N}{\omega}\sin(\omega t)\right]}}
= \frac{\tanh\left(\frac{\alpha_N}{\omega} \sin(\omega t)\right) + M_{x0}}
{1 + M_{x0} \tanh\left(\frac{\alpha_N}{\omega}\sin(\omega t)\right)},
\nonumber \\
M_{y}(t) &=& \sqrt{1-M_{x}^{2}(t)} \, \,
\sin\left[\frac{\omega_L}{\omega}\sin(\omega t) + \delta_0\right],
\nonumber \\
M_{z}(t) &=& \sqrt{1-M_{x}^{2}(t)} \, \,
\cos\left[\frac{\omega_L}{\omega}\sin(\omega t) + \delta_0\right],
\end{eqnarray}
where, as before, $M_{x0}$ and $\delta_0$ are determined by the initial 
conditions: $M_{x0} = M_x(0)$ and $\delta_0 = \tan^{-1}(M_{y}(0)/M_{z}(0))$.  
The interpretation of equation (\ref{osc_sol_relax}) is that, like in 
absence of relaxation, the unit vector representing the magnetization 
is precessing  around the $x$ axis with a time-dependent angular velocity, 
but in this case the angle $\beta$ between the {\bf M} vector and the $x$ 
axis changes as $\beta=\cos^{-1}(M_{x}(t))$. At time $t$, the projection 
of {\bf M} on the $yz$ plane has a length of $\sin(\beta)$ and is at an 
angle $\delta(t) = \tan^{-1}(M_{y}(t)/M_{z}(t)) = [(\omega_L /\omega) 
\sin(\omega t) + \delta_0$] from the $z$ axis. The angular velocity of the 
precession and the projected length of the unit vector {\bf M} are changing 
periodically. If $(\alpha_N/\omega)<<1$ and $(\omega_L/\omega)>>1$ the 
precessing unit vector approaches slowly the $x$ axis, completing many 
cycles before it slows down and changes sign. Note that the first condition 
can be fulfilled both at very weak and very strong damping, since 
$\alpha_N \propto (1+\alpha^2)^{-1}$. For $(\alpha_N/\omega)>>1$ the 
precessing unit vector approaches quickly the direction of the driving 
field, it is very close to being aligned with the $x$ axis, alternating 
between the positive and negative directions. For vanishingly weak or 
infinitely strong damping, that is in the $\alpha_N \to 0$ limit, the 
solution reduces to equation (\ref{osc_sol_norelax}). For the static 
limit, i.e., $\omega \to 0$ the general solution (\ref{osc_sol_relax}) 
reduces to 
\begin{eqnarray}
\label{stat_sol_relax}
M_{x}(t) &=& \frac{(M_{x0}-1)+(M_{x0}+1)
\exp{\left[2\alpha_N t\right]}} 
{(1-M_{x0})+(M_{x0}+1)
\exp{\left[2\alpha_N t\right]}}
= \frac{\tanh\left(\alpha_N t \right) + M_{x0}}
{1 + M_{x0} \tanh\left(\alpha_N t \right)},
\nonumber \\
M_{y}(t) &=& \sqrt{1-M_{x}^{2}(t)} \, \,
\sin\left[\omega_L t + \delta_0\right],
\nonumber \\
M_{z}(t) &=& \sqrt{1-M_{x}^{2}(t)} \, \,
\cos\left[\omega_L t + \delta_0\right].
\end{eqnarray}
In this case the {\bf M} vector is precessing with a time-independent 
angular velocity $\omega_L$ and converges towards the $+x$ direction with 
a time constant  $\sim 1/\alpha_N$. In practice, for $H\approx 10^{4}$ 
Am$^{-1}$, we find that $\alpha_N = \mu_0 \gamma_0 H \alpha /(1+\alpha^2)
\approx \alpha/(1+\alpha^2)\cdot 10^{9}$ s$^{-1}$, so that even for weak 
damping, e.g.,  $\alpha= 0.001$, the convergence is very fast, $M_y$ and 
$M_z$ vanish within a few microseconds. This implies that for frequencies 
much lower than 1 MHz we have $M_x(t)\cong 1$.

%------------------------------------------------------------
\subsection{Specific loss power}
%------------------------------------------------------------

The results given above enable us to calculate the energy loss for a 
single particle. The energy dissipated in a single cycle can be 
calculated as 
\begin{eqnarray}
\label{def_loss}
E = \mu_0 \int_{0}^{2\pi/\omega} \mr{d}t \, \, 
\left({\bf H} \cdot \frac{d{\bf m}}{dt} \right)
= \mu_0 m_S \int_{0}^{2\pi/\omega} \mr{d}t \, \, 
\left({\bf H} \cdot \frac{d{\bf M}}{dt} \right).
\end{eqnarray}
Using the expression (\ref{osc_field}) for ${\bf H}$ and the LLG
equation (\ref{osc_eq}) and its solution Eq.(\ref{osc_sol_relax}),
the energy dissipated in a single cycle, 
\begin{eqnarray}
\label{exact_loss_osc}
E = \mu_0 m_S \alpha_N H \int_{0}^{2\pi/\omega} \mr{d}t \, \,
\, [\cos(\omega t)]^2 \, \, [ 1- M_x^2(t) ],
%= \mu_0 m_S H \alpha' \int_{0}^{2\pi/\omega} \mr{d}t \, \,
%\, [\cos(\omega t)]^2 \, \, [ 1- M_x^2(t) ]
\end{eqnarray}
can be calculated numerically. The loss per cycle depends on the 
initial conditions, it has a maximum when $M_{x0}=0$.

Analytic results are available in the limit of $(\alpha_N/\omega) <<1$ 
where $M_x(t)$ can be expanded in powers of $\alpha_N/\omega$. 
As in the initial state, in a ferrofluid, the magnetization vectors are 
randomly distributed, the relevant quantity will be the average of the 
loss power (\ref{exact_loss_osc})
over this distribution. To lowest order in $\alpha_N/\omega$
this gives
\begin{eqnarray}
\label{loss_osc_average_weak}
E_{\mr{average}} \approx
\frac{2}{3}\mu_0 \pi m_S H
\frac{\alpha_N}{\omega}.
\end{eqnarray}
%
%that is, the maximum loss is reduced by only a factor $2/3$. 

At frequencies much lower than 1 MHz ($\alpha_N/\omega >>1$), the loss per 
cycle becomes independent of the initial conditions and frequency. 
%[unless $\vert M_{x0}\vert < 1-\ord{\exp(-\alpha_N/\omega)}$] 
In this limit it equals 
\begin{eqnarray}
\label{loss_osc_average_strong}
E_{\mr{average}} \approx E = 4 \mu_0 m_S H.
\end{eqnarray}
%
%In the
%opposite limit, for $\alpha_N/\omega >>1$ the loss per cycle tends to the 
%constant (\ref{loss_osc_strong}) which is independent of the initial 
%conditions, except $M_{x0}=1$ where it is zero. 

%-------------------------------------------------------------- 
% Circularly polarized applied field
%--------------------------------------------------------------
\section{Circularly polarized applied field} 
\label{circ} 
Here, we present analytic solutions of the LLG equations for an immobile
single-domain magnetic particle under circularly polarized, i.e. rotating
applied magnetic field. Our exact analytic solutions are only valid for
specific initial conditions. However, they also give the correct answer if one
is interested in the behaviour in the steady state that sets in after a short
relaxation process. The heating loss per cycle of a single particle in this
state will be calculated to be compared in the next section to that obtained
for the linearly polarized field.

The applied field is assumed to rotate in the $xy$ plane with an angular 
frequency $\omega$
\begin{eqnarray}
\label{rot_field}
{\bf H} &=& 
\frac{\omega_L}{\gamma'} \, \, (\cos(\omega t), \sin(\omega t), 0),
%\nonumber \\
%{\bf H} &=& 
%\frac{\alpha_N}{\alpha'} \, \, (\cos(\omega t),\sin(\omega t) , 0),
\end{eqnarray}
We also introduce the angular velocity vector ${\bf \omega}$, which in 
this case is perpendicular the the $xy$ plane.

%--------------------------------------------------------------
\subsection{Free precession} 
%--------------------------------------------------------------

The LLG equations for the Cartesian components of the magnetization are
\begin{eqnarray}
\label{rot_nonrelax_eq}
\frac{\mr{d}}{\mr{d}t} M_x &=& 
-\omega_L \, M_z \, \sin(\omega t),
\nonumber \\
\frac{\mr{d}}{\mr{d}t} M_y &=& 
\omega_L \, M_z \, \cos(\omega t),
\nonumber \\
\frac{\mr{d}}{\mr{d}t} M_z &=& 
\omega_L \,\left(M_x \sin(\omega t) - M_y \cos(\omega t) \right).
\end{eqnarray}
It is easy to find a special solution of these coupled equations for 
the case when $d(M_z)/dt =0$, i.e., $M_z$ is a time-independent constant: 
\begin{eqnarray}
\label{rot_sol_spec}
M_{x}(t) &=&  \frac{\omega_L}{\sqrt{\omega^2 + \omega_L^2}} 
\cos(\omega t),
\nonumber \\
M_{y}(t) &=&   \frac{\omega_L}{\sqrt{\omega^2 + \omega_L^2}} 
\sin(\omega t),
\nonumber \\
M_{z}(t) &=& \frac{\omega}{\sqrt{\omega^2 + \omega_L^2}}.
\end{eqnarray}
This solution is valid only for a special initial condition, {\em viz.},
$M_x(0)/M_z(0) = \omega_L/\omega$, $M_y(0)=0$. For any other 
initial values, one has to find the general solution of equation 
(\ref{rot_nonrelax_eq}). This can be done by means of an appropriate 
rotation into a coordinate system in which the LLG equations have the 
simplest possible solution: a time-independent magnetization vector. 
The transformation is done in three stages. The first rotation,
\begin{align} 
\label{rot1}
\underline{\underline{\mr{O}_1}} = 
\left(\begin{array}{ccc} 
+\cos(\omega t) &\hspace*{0.2cm} +\sin(\omega t) &\hspace*{0.2cm} 0 
\\
-\sin(\omega t) &\hspace*{0.2cm} +\cos(\omega t) &\hspace*{0.2cm} 0
\\
0               &\hspace*{0.2cm} 0              &\hspace*{0.2cm} 1 
\end{array} \right)
\end{align}
transforms the vector equation (\ref{rot_nonrelax_eq}) into a coordinate 
system, which rotates around the $z$ axis with the applied magnetic field. 
The transformed $z$ axis points then in the direction of the angular 
velocity vector ${\bf \omega}$. The second rotation, 
\begin{align} 
\label{rot2}
\underline{\underline{\mr{O}_2}} = 
\left(\begin{array}{ccc} 
+\cos(\Theta) &\hspace*{0.2cm} 0 &\hspace*{0.2cm} -\sin(\Theta)  
\\
0             &\hspace*{0.2cm} 1 &\hspace*{0.2cm} 0 
\\
+\sin(\Theta) &\hspace*{0.2cm} 0 &\hspace*{0.2cm} +\cos(\Theta) 
\end{array} \right)
\end{align}
tilts the $z$ axis into the direction of the angular velocity vector
${\bf \Omega} = {\bf \omega} + {\bf \omega}_L$. As ${\bf \omega}_L$ is
aligned with ${\bf H}$, it is perpendicular to ${\bf \omega}$, so that
$\vert{\bf \Omega}\vert = \Omega = \sqrt{\omega^2 + \omega_L^2}$
and $\cos(\Theta)=\omega/\Omega$ and $\sin(\Theta)=\omega_L/\Omega$.
Finally, the last rotation 
\begin{align} 
\label{rot3}
\underline{\underline{\mr{O}_3}} = 
\left(\begin{array}{ccc} 
+\cos(\Omega t) &\hspace*{0.2cm} -\sin(\Omega t) &\hspace*{0.2cm} 0 
\\
+\sin(\Omega t) &\hspace*{0.2cm} +\cos(\Omega t) &\hspace*{0.2cm} 0
\\
0               &\hspace*{0.2cm} 0              &\hspace*{0.2cm} 1 
\end{array} \right)
\end{align}
is again a transformation into a rotating coordinate system. Similar 
rotations have been discussed in referencies \cite{SunWang} and \cite{Denisov}. 
In the new rotating coordinate system the transformed magnetization vector 
$(\underline{\underline{\mr{O}_3}}\cdot\underline{\underline{\mr{O}_2}}
\cdot\underline{\underline{\mr{O}_1}}{\bf M}) = (M_{\xi},M_{\eta},M_{\zeta})$ 
is found to be time-independent. The inverse transformation 
$(\underline{\underline{\mr{O}^{-1}_1}}
\cdot\underline{\underline{\mr{O}^{-1}_2}}
\cdot\underline{\underline{\mr{O}^{-1}_3}})$ gives then  
the general solution of the 
LLG equation for circularly polarized applied field in the limit of 
vanishing N\'eel relaxation, which has the following 
form
\begin{eqnarray}
\label{rot_sol_nonrelax}
M_{x}(t) &=&  \frac{\omega}{\Omega} \cos(\omega t)
\left[M_{\xi} \cos(\Omega t) + M_{\eta} \sin(\Omega t)\right]
+ \sin(\omega t) \left[M_{\xi} \sin(\Omega t) - M_{\eta} \cos(\Omega t)\right]
+ \frac{\omega_L}{\Omega} \cos(\omega t) \sqrt{1-M_{\xi}^2 -M_{\eta}^2},
\nonumber \\
M_{y}(t)  &=&   \frac{\omega}{\Omega} \sin(\omega t)
\left[M_{\xi} \cos(\Omega t) + M_{\eta} \sin(\Omega t)\right]
- \cos(\omega t) \left[M_{\xi} \sin(\Omega t) - M_{\eta} \cos(\Omega t)\right]
+ \frac{\omega_L}{\Omega} \sin(\omega t) \sqrt{1-M_{\xi}^2 -M_{\eta}^2},
\nonumber \\
M_{z}(t) &=& -\frac{\omega_L}{\Omega}
\left[M_{\xi} \cos(\Omega t) + M_{\eta} \sin(\Omega t)\right]
+\frac{\omega}{\Omega}\sqrt{1-M_{\xi}^2 -M_{\eta}^2},
\end{eqnarray}
where $M_{\xi}$ and $M_{\eta}$ are the Cartesian components of the 
magnetization vector in the rotated coordinate system. They are determined 
by the initial conditions as
\begin{equation}
\label{init_ax0_ay0}
M_{\xi} = \frac{\omega M_{x0}
-\omega_L \sqrt{1-(M_{x0})^2 -(M_{y0})^2}}{\Omega},
\hskip 0.5cm
M_{\eta} = M_{y0}.
\end{equation}
($M_{\xi}$ and $M_{\eta}$ are time-independent.) For non-vanishing 
relaxation, i.e. for $\alpha_N \neq 0$, $M_{\xi}=M_{\xi}(t)$ and 
$M_{\eta}=M_{\eta}(t)$ become time-dependent. It is easily verified that
$M_{\xi} = M_{\eta} = 0$ implies the initial conditions for the validity
of the special solution (\ref{rot_sol_spec}) and indeed equation 
(\ref{rot_sol_nonrelax}) reduces to equation (\ref{rot_sol_spec}) in this
case. In the limit of $\omega \to 0$, equation (\ref{rot_sol_nonrelax}) 
recovers the solution (\ref{stat_sol_norelax}) which is obtained for static 
applied field with $\alpha_N \to0$.

%--------------------------------------------------------------
\subsection{Damped precession} 
%--------------------------------------------------------------

The LLG equations for the Cartesian components of the magnetization of 
a single magnetic nanoparticle in the presence of N\'eel relaxation, under 
circularly polarized applied field are 
\begin{eqnarray}
\label{rot_relax_eq}
\frac{\mr{d}}{\mr{d}t} M_x &=& -\omega_L \, M_z \, \sin(\omega t) 
+ \alpha_N 
\left[-M_x M_y \sin(\omega t) + (M_y^2 + M_z^2) \cos(\omega t)\right],
\nonumber \\
\frac{\mr{d}}{\mr{d}t} M_y &=&  \omega_L \, M_z \, \cos(\omega t)
+ \alpha_N 
\left[-M_x M_y \cos(\omega t) + (M_x^2 + M_z^2) \sin(\omega t)\right],
\nonumber \\
\frac{\mr{d}}{\mr{d}t} M_z &=& 
\omega_L \,\left[M_x \sin(\omega t) - M_y \cos(\omega t) \right]
- \alpha_N 
\left[ M_x M_z \cos(\omega t) + M_y M_z \sin(\omega t)\right].
\end{eqnarray}
The general solution for arbitrary initial conditions can only be 
obtained numerically. It can be verified though that the LLG equations 
(\ref{rot_relax_eq}) are satisfied by the intuitively guessed solution
\begin{eqnarray}
\label{rot_sol_relax_spec}
M_{x}(t) &=& M_{xy} \cos(\omega t -\varphi) 
= M_{x0}^{\mr{spec}} \cos(\omega t) - M_{y0}^{\mr{spec}} \sin(\omega t),
\nonumber \\
M_{y}(t) &=& M_{xy} \sin(\omega t -\varphi)
= M_{x0}^{\mr{spec}} \sin(\omega t) + M_{y0}^{\mr{spec}} \cos(\omega t),
\nonumber \\
M_{z}(t) &=& \sqrt{1-(M_{x0}^{\mr{spec}})^2 -(M_{y0}^{\mr{spec}})^2}.
\end{eqnarray}
where $M_{xy}$ is the projection of the magnetization vector onto 
the $xy$ plane, $M_{x0}^{\mr{spec}} = M_{xy} \cos(\varphi)$, 
$M_{y0}^{\mr{spec}} = -M_{xy} \sin(\varphi)$
and 
\begin{eqnarray}
\label{mx0_my0}
M_{x0}^{\mr{spec}} &=& \sqrt{\frac{\alpha_N^2 -\omega_L^2 -\omega^2 + 
\sqrt{4\alpha_N^2 \omega_L^2 
+ (\alpha_N^2 -\omega_L^2 -\omega^2)^2}}{2\alpha_N^2}},
\nonumber \\
M_{y0}^{\mr{spec}} &=& - \frac{\alpha_N^2 +\omega_L^2 +\omega^2 - 
\sqrt{4\alpha_N^2 \omega_L^2 
+ (\alpha_N^2 -\omega_L^2 -\omega^2)^2}}{2\omega\alpha_N}.
\end{eqnarray}
Equation (\ref{rot_sol_relax_spec}) describes a precession around the 
$z$ axis with a phase shift relative to the rotation of the magnetic 
field. However, there are conditions for the initial state: the phase 
shift is strictly determined by the tilt of the magnetization vector 
with respect to the $z$ axis. The tilt is given by $\theta = \cos^{-1}
\left(\sqrt{1-(M_{x0}^{\mr{spec}})^2-(M_{y0}^{\mr{spec}})^2}\right)$, 
and the phase shift by 
$\varphi = \tan^{-1}(-M_{y0}^{\mr{spec}}/M_{x0}^{\mr{spec}})$. The 
relation between $\theta$ and $\varphi$ is thus implicit in equation 
(\ref{mx0_my0}).

We have verified with countless numerical calculations that any 
other solution of the LLG equation (\ref{rot_relax_eq}) tends to 
equation (\ref{rot_sol_relax_spec}) after a rather small transient 
time-interval. For the ensuing steady state the solutions of the LLG 
equation can be well approximated by equation (\ref{rot_sol_relax_spec}) . 
Similar asymptotic solution of the general LLG equation for rotating 
applied field has been discussed recently by Sun and Wang \cite{SunWang}. 

\begin{figure}[htb]  
\begin{center}  
\begin{minipage}{14cm} 
\begin{center} 
\epsfig{file=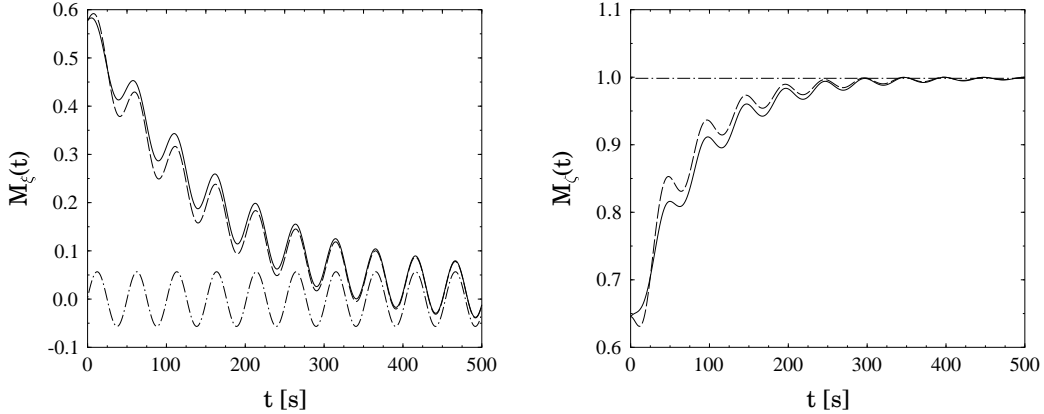,width=1.0\linewidth}  
\caption{\label{fig_approx} Numerical (full line), approximate 
(dashed line), see Eq.~(\ref{rot_sol_relax_approx_a}) and the special 
(dashed-dotted line), see Eq.~(\ref{rot_sol_relax_spec_a}) solutions 
of the LLG equations for the rotated coordinates $M_{\xi}(t)$ and 
$M_{\zeta}(t)$ of the magnetization are compared for 
$\omega=\omega_L = 0.088$ s$^{-1}$, $\alpha =0.14$.  
}  
\end{center} 
\end{minipage} 
\end{center}  
\end{figure}

To understand the coupled dynamics of relaxation, precession and 
rotation it is useful to rewrite the LLG equation with N\'eel 
relaxation, equation (\ref{rot_relax_eq}), and the special solution 
(\ref{rot_sol_relax_spec}) in the rotating coordinate system defined 
in connection with the relaxation-free case. The transformed LLG 
equations read 
\begin{eqnarray}
\label{rot_relax_eq_a}
\frac{\mr{d}}{\mr{d}t} M_{\xi} &=& \alpha_N \left[
- M_{\xi} M_{\eta} \frac{\omega}{\Omega}\sin(\Omega t) 
- M_{\xi} M_{\zeta} \frac{\omega_L}{\Omega}
+ (1 - M_{\xi}^2) \frac{\omega}{\Omega} \cos(\Omega t)
\right],
\nonumber \\
\frac{\mr{d}}{\mr{d}t} M_{\eta} &=& \alpha_N \left[
- M_{\eta} M_{\xi} \frac{\omega}{\Omega} \cos(\Omega t) 
- M_{\eta} M_{\zeta} \frac{\omega_L}{\Omega}
+ (1 - M_{\eta}^2) \frac{\omega}{\Omega} \sin(\Omega t)
\right],
\nonumber \\
\frac{\mr{d}}{\mr{d}t} M_{\zeta} &=&  \alpha_N \left[
- M_{\zeta} M_{\xi} \frac{\omega}{\Omega} \cos(\Omega t) 
- M_{\zeta} M_{\xi} \frac{\omega}{\Omega} \sin(\Omega t)
+ (1 - M_{\zeta}^2) \frac{\omega_L}{\Omega}
\right],
\end{eqnarray}
where $(M_{\xi}, M_{\eta}, M_{\zeta})$ represents the magnetization 
vector in the rotated coordinate system. As {\bf M} remains a unit 
vector in its transformed form, the LLG equations for the components 
$M_{\xi}$, $M_{\eta}$ and $M_{\zeta}$ are not independent. The special 
solution (\ref{rot_sol_relax_spec}) in terms of the rotating 
coordinates has the following form
\begin{eqnarray}
\label{rot_sol_relax_spec_a}
M_{\xi}(t) &=& M_{\xi\eta} \cos(\Omega t -\varphi_a) 
= M_{\xi 0}^{\mr{spec}} \cos(\Omega t) - M_{\eta 0}^{\mr{spec}} 
\sin(\Omega t),
\nonumber \\
M_{\eta}(t) &=& M_{\xi\eta} \sin(\Omega t -\varphi_a)
= M_{\xi 0}^{\mr{spec}} \sin(\Omega t) + M_{\eta 0}^{\mr{spec}} 
\cos(\Omega t),
\nonumber \\
M_{\zeta}(t) &=& \sqrt{1-(M_{\xi 0}^{\mr{spec}})^2 -
(M_{\eta 0}^{\mr{spec}})^2},
\end{eqnarray}
where  $M_{\xi 0}^{\mr{spec}} = M_{\xi\eta} \cos(\varphi_a)$, 
$M_{\eta 0}^{\mr{spec}} = -M_{\xi\eta} \sin(\varphi_a)$ and similarly 
to equation (\ref{init_ax0_ay0}), the time-independent parameters 
are 
\begin{equation}
\label{axo_ayo}
M_{\xi 0}^{\mr{spec}} = 
\frac{\omega M_{x0}^{\mr{spec}}
-\omega_L \sqrt{1-(M_{x0}^{\mr{spec}})^2 -(M_{y0}^{\mr{spec}})^2}}{\Omega},
\hskip 0.5cm
M_{\eta 0}^{\mr{spec}} = M_{y0}^{\mr{spec}}. 
\end{equation}
The advantage of the rotating coordinate system is that the time 
dependence that stands for the rotation is embedded inside the 
relaxation term, showing that the motion is dominated by the trend 
towards the stationary solution by a characteristic time constant 
$\sim 1/\alpha_N$. Therefore, it represents a good framework to look 
for the general solution of the LLG equation or, alternatively, one can 
try to find approximate solutions of the LLG equation. For example 
\begin{eqnarray}
\label{rot_sol_relax_approx_a}
M_{\xi}(t) &=&  
M_{\xi 0}^{\mr{spec}} \cos(\Omega t) - M_{\eta 0}^{\mr{spec}} \sin(\Omega t)
+ (M_{\xi 0} - M_{\xi 0}^{\mr{spec}}) \, \, e^{-\frac{\alpha_N t}{\sqrt{2}}},
\nonumber \\
M_{\eta}(t) &=& 
M_{\xi 0}^{\mr{spec}} \sin(\Omega t) + M_{\eta 0}^{\mr{spec}} \cos(\Omega t)
+ (M_{\eta 0} - M_{\eta 0}^{\mr{spec}}) \, \, e^{-\frac{\alpha_N t}{\sqrt{2}}},
\nonumber \\
M_{\zeta}(t) &=& 
\sqrt{1-[M_{\xi}(t)]^2 -[M_{\eta}(t)]^2},
\end{eqnarray}
well approximates the numerical solution of the LLG equation. 
The exponential decay of the last contributions to the in-plane 
components shows that the general solution converges quickly to the 
special solution. This is demonstrated in figure \ref{fig_approx}, 
for $\alpha_N = 0.01 s^{-1}$, ($\alpha = 0.14$) and 
$\omega = \omega_L = 0.088$ s$^{-1}$, where the numerical and 
approximate solutions of the LLG equation are compared. 
The other consequence of the short transient process is, that 
irrespective of the initial conditions, the magnetization vector lies 
in the plane of the rotation, where it follows the driving field with 
a constant phase slip.
%
% Fig. 1  
%  

We note that both in the limits of weak and strong relaxation the 
phase slip $\varphi$ goes to zero. This is seen in the values of the 
parameters in equation (\ref{mx0_my0}) of the special solution 
(\ref{rot_sol_relax_spec}) in the appropriate limits: 
\begin{equation}
\label{mxo_myo_weak}
M_{x0}^{\mr{spec}}(\alpha_N \to 0) 
= \frac{\omega_L}{\sqrt{\omega^2 + \omega_L^2}},  
\hskip 0.5cm 
M_{y0}^{\mr{spec}}(\alpha_N \to 0) = 0,
\end{equation}
and
\begin{equation}
\label{mxo_myo_strong}
M_{x0}^{\mr{spec}}(\alpha_N\to 10^8 s^{-1}) \approx 1,  
\hskip 0.5cm 
M_{y0}^{\mr{spec}}(\alpha_N\to 10^8 s^{-1}) \approx 0,
\end{equation}
as in the limit of strong relaxation $\alpha_N \approx 10^8$ s$^{-1}$ 
values can be achieved. In general, the phase slip will depend on 
$\omega_L(H)$ and $\omega$ according to the relation $\varphi
= \tan^{-1}(-M_{y0}^{\mr{spec}}/M_{x0}^{\mr{spec}})$.

%------------------------------------------------------------
\subsection{Specific loss power}
%------------------------------------------------------------

To calculate the energy loss per cycle for a single particle 
we use again equation (\ref{def_loss}). The special solution 
(\ref{rot_sol_relax_spec}) can be used to assess the merits of 
circular polarization, because in hyperthermia the field is applied 
for a long time ($\approx 10^3$ s) compared to the duration of the 
initial transient process ($\approx 10^{-6}$ s). Using this special 
solution, the dot product in equation (\ref{def_loss}) turns out to 
be integrable and the energy loss per cycle is found to be
\begin{eqnarray}
E = \mu_0 2\pi m_S H (-M^{\mr{spec}}_{y0}).
\end{eqnarray}
The weak-relaxation (or high-frequency) limit, 
$\alpha_N << \omega,\omega_L$, can be studied by means of a Taylor 
expansion of $(-M^{\mr{spec}}_{y0})$, 
in powers of $\alpha^2_N/(\omega_L^2 + \omega^2)$. The result,
\begin{equation}
\label{loss_rot_max}
E \approx  2 \pi \mu_0 m_S H
\left[\frac{\omega^2}{\omega_L^2 + \omega^2}
\left(\frac{\alpha_N}{\omega}\right) 
-\frac{\omega^4 \omega_L^2}{(\omega_L^2 + \omega^2)^3}
\left(\frac{\alpha_N}{\omega}\right)^3 \right] 
\end{equation}
shows that there is no dissipation without relaxation. Making use 
of the relation $\alpha_N = \alpha \omega_L$ 
this expression can be written in a form that highlights the scaling 
of the variables:
\begin{equation}
\label{loss_rot_weak}
E \approx  2 \pi \mu_0 m_S H \frac{\omega}{\omega_L}
\left[\left(\frac{\alpha}{1+(\omega/\omega_L)^2}\right)
-\left(\frac{\alpha}{1+(\omega/\omega_L)^2}\right)^3
\right].
\end{equation}
The loss per cycle for strong relaxation (or low-frequency), 
$\alpha_N >> \omega,\omega_L$, shows similar scaling relations
\begin{equation}
E \approx  2 \pi \mu_0 m_S H  
\left[\frac{\alpha_N^2}{\omega_L^2 + \alpha_N^2}
\left(\frac{\omega}{\alpha_N}\right) 
- \frac{\alpha_N^4 \omega_L^2}{(\omega_L^2 + \alpha_N^2)^3}
\left(\frac{\omega}{\alpha_N}\right)^3\right],
\end{equation}
and
\begin{equation}
\label{loss_rot_strong}
E \approx  2 \pi \mu_0 m_S H \frac{\omega}{\omega_L}
\left[\left(\frac{\alpha}{1+(\alpha)^2}\right)
-\left(\frac{\alpha}{1+(\alpha)^2}\right)^3
\right].
\end{equation}
We emphasize that the distinction between $\omega_L$ and the 'bare' 
Larmor frequency $\omega_{L0} = \mu_0 \gamma_0 H$ cannot be ignored 
in the strong-relaxation limit. In fact, the definition 
$\alpha_N = \omega_{L0} \alpha/(1+\alpha^2)$ makes the limit 
$\alpha_N \to \infty$ meaningless, because the largest value of  
$\alpha_N$ is $\omega_{L0}/2$, which is realized at $\alpha = 1$.

%--------------------------------------------------------------
% Conclusion and Summary 
%--------------------------------------------------------------
\section{Conclusion and Summary} 
\label{sum} 
 
We have reported numerical and analytical results on the response of 
an immobilised single-domain magnetic particle under linearly and 
circularly polarized magnetic fields. Comparison of the dynamics in 
the presence and absence of N\'eel relaxation enables an interpretation 
of the results in terms of an interplay between Larmor preccesion and 
relaxation towards energy minima. 

%
% Fig. 2
%  
\begin{figure}[htb]  
\begin{center}  
\begin{minipage}{14cm} 
\begin{center}   
\epsfig{file=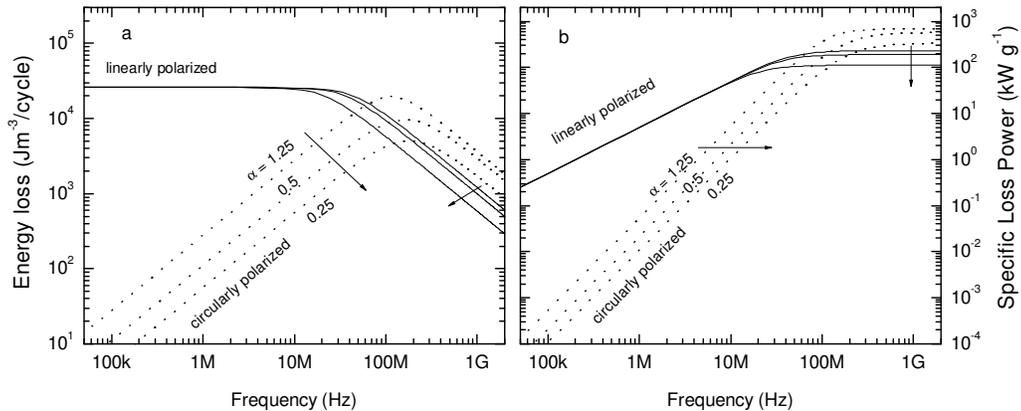,width=1.0\linewidth}  
\caption{\label{fig_loss} 
Energy loss per cycle (a) and specific loss power (b) for linearly 
and circularly polarised fields for magnetite as a function of 
driving field frequency for different $\alpha$ damping parameters. 
Arrows point towards the direction of decreasing $\alpha$. Driving 
field amplitude H = $10^4$ A/m.  
}  
\end{center} 
\end{minipage} 
\end{center}  
\end{figure} 

In figure 2a the resulting energy loss per cycle is depicted as a 
function of frequency and the dimensionless damping parameter $\alpha$. 
Regarding the merits of linear or circular polarization of the 
magnetic field, we find opposite preferences for frequencies below 
and above the Larmor frequency. In the former case, which is relevant 
to hyperthermia, the energy absorbed by the magnet from linearly 
polarised field exceeds the absorption from rotating field by 
orders of magnitude. Apparently the relaxation towards a steady 
state with a small phase shift between the rotating {\bf H} and 
{\bf m} fields ensures a smooth process with very small losses. 
This cannot take place in a linear field, where the orientation of 
{\bf H} changes abruptly twice in a cycle and relaxation towards the 
new energy minimum has to be repeated. In the case of high frequencies, 
which is not lacking technological relevance since the recent interest 
in materials with high microwave absorption \cite{LiGeEtAl2008}, the 
relaxation is slow compared to the rate of change of the magnetic field 
and the loss per cycle becomes frequency dependent.

The tent-shaped curves in figure 2a reflect the relevance of the 
limiting cases worked out in sections 3.3 and 4.3. The low-frequency 
behaviour, which seems to prevail through many orders of magnitude in 
frequency, reflects equation (\ref{stat_sol_relax}) for the case of 
linear polarization and the first term in equation (\ref{loss_rot_weak}) 
for rotating field. Equation (\ref{stat_sol_relax}) 
describes a relaxation process, which takes place after each change 
of sign of $H_x$, apparently at the same cost in energy as long as the 
period of the oscillation of the field is much larger than the 
relaxation time. The loss per cycle is then always the same and we 
see no dependence on the damping constant either, except that for 
weaker damping the deviation from the constant value begins at a 
lower frequency. Equation (\ref{loss_rot_weak}) shows clearly that for 
frequencies much lower than the Larmor frequency the loss is always 
proportional to the frequency, the slope scaling with the dimensionless 
damping factor $\alpha$ (in the log-log plot this translates to straight 
lines with the same slope, shifted by distances scaling with $\alpha$). In 
contrast, equations (\ref{loss_osc_average_weak}) and (\ref{loss_rot_max}) 
show that, in both cases, at high frequencies the loss is proportional 
to $\alpha_N/\omega$, which explains why all curves run 
parallel, with he same slope, and the dependence of their position 
on $\alpha$ is not monotonic.

Figure 2b shows the same data in practical units, that is, it shows 
the proper specific absorption power in W/g for a single-domain 
particle of magnetite. At frequencies relevant to hyperthermia 
($\sim 10^5$ Hz) with linearly polarized field, respectable losses of 
kW/g order can be achieved, while the power absorbed from circularly 
polarized field is two to three orders of magnitude lower. Our 
final conclusion is obvious: if N\'eel relaxation in isotropic sample 
is the dominant mechanism, the technical complications of generating 
a circularly polarised field in difficult geometry need not be 
considered.

%-------------------------------------------------------------
% Acknowledgments 
%-------------------------------------------------------------
\section*{Acknowledgments} 
 
The authors acknowledge support from the Hungarian
National Office for Research and Technology NKFP-5/006/2005, 
Contract OM-00077/2005.

%--------------------------------------------------------------

\end{document}